\documentclass[twocolumn,prl,showpacs,amsfonts,amsmath,assymb,eufrak]{revtex4}

\usepackage{epsfig}
\def\eq#1\en{\begin{equation}#1\end{equation}}  
\def\eqa#1\ena{\begin{align}#1\end{align}}
\def\eqg#1\eng{\begin{gather}#1\end{gather}}
\newcommand{\lb}[1]{\label{e:#1}}
\newcommand{\rlb}[1]{\eqref{e:#1}} 
\newcommand{\nl}{\notag\\}

\newcommand{\abs}[1]{\left|#1\right|}

\newcommand{\bkt}[1]{\left\langle#1\right\rangle}
\newcommand{\sbkt}[1]{\langle#1\rangle}

\newcommand{\sumtwo}[2]%
{\mathop{\sum_{#1}}_{#2}}
\newcommand{\sumthree}[3]%
{\mathop{\mathop{\sum_{#1}}_{#2}}_{#3}}
\newcommand{\sumfour}[4]%
{\mathop{\mathop{\mathop{\sum_{#1}}_{#2}}_{#3}}_{#4}} 
\newcommand{\prodtwo}[2]%
{\mathop{\prod_{#1}}_{#2}}
\newcommand{\mintwo}[2]%
{\mathop{\min_{#1}}_{#2}}
\newcommand{\maxtwo}[2]%
{\mathop{\max_{#1}}_{#2}}
\newcommand{\maxthree}[3]%
{\mathop{\mathop{\max_{#1}}_{#2}}_{#3}}
\newcommand{\limtwo}[2]%
{\mathop{\lim_{#1}}_{#2}}
\newcommand{\suptwo}[2]%
{\mathop{\sup_{#1}}_{#2}}
\newcommand{\supthree}[3]%
{\mathop{\mathop{\sup_{#1}}_{#2}}_{#3}}
\newcommand{\supfour}[4]%
{\mathop{\mathop{\mathop{\sup_{#1}}_{#2}}_{#3}}_{#4}} 
\newcommand{\inftwo}[2]%
{\mathop{\inf_{#1}}_{#2}}
\newcommand{\infthree}[3]%
{\mathop{\mathop{\inf_{#1}}_{#2}}_{#3}}
\newcommand{\inffour}[4]%
{\mathop{\mathop{\mathop{\inf_{#1}}_{#2}}_{#3}}_{#4}} 
\newcommand\calD{{\cal D}}

\newcommand\calH{{\cal H}}

\newcommand\calO{{\cal O}}

\newcommand{\Di}{\mathit{\Delta}}

\newcommand{\DU}{\Di U}
\newcommand{\al}{\alpha}
\newcommand{\ph}{\varphi}
\newcommand{\hH}{\hat{H}}
\newcommand{\hP}{\hat{P}}
\newcommand{\ket}[1]{|#1\rangle}
\newcommand{\kpj}{\ket{\psi_j}}
\newcommand{\kpz}{\ket{\ph(0)}}
\newcommand{\kx}{\ket{\xi^{(\nu)}}}
\newcommand{\bn}{\beta^{(\nu)}}

\newcommand{\sumj}{\sum_{j=1}^D}
\newcommand{\sumn}{\sum_{\nu=1}^d}
\newcommand{\Heq}{\calH_\mathrm{eq}}
\newcommand{\Hneq}{\calH_\mathrm{neq}}
\newcommand{\para}[1]{{\em #1}\/.---}
\newcommand{\midskip}{\vspace{3pt}}
\begin{document}
\title{On the time scales in the approach to equilibrium of macroscopic quantum systems}

\author{Sheldon Goldstein${}^1$, Takashi Hara${}^2$, and Hal Tasaki${}^3$}
\affiliation{
${}^1$%
Departments of Mathematics and Physics, Rutgers University, 110 Frelinghuysen Road, Piscataway, NJ 08854-8019, USA
\\
${}^2$%
Faculty of Mathematics,
Kyushu University,
Moto-oka, Nishi-ku,
Fukuoka 819-0395,
Japan
\\
${}^3$%
Department of Physics, Gakushuin University, 
Mejiro, Toshima-ku, Tokyo 171-8588, Japan}

\date{\today}

%%%%%%%%%%%%%%%%%
\begin{abstract}
We prove two theorems concerning the time evolution in general isolated quantum systems.
The theorems are relevant to the issue of the time scale in the approach to equilibrium.
The first theorem shows that there can be pathological situations in which the relaxation takes an extraordinarily long time, while the second theorem shows that one can always choose an equilibrium subspace the relaxation to which requires only a short time for any initial state.
\end{abstract}

% 05.30.-d 	Quantum statistical mechanics 
%05.70.-a 	Thermodynamics
% 03.65.Yz 	Decoherence; open systems; quantum statistical methods

\maketitle
%%%%%%%%%%%%%%%%%%%%%%%%%%%%%%%%%%%%%
%%%%%%%%%%%%%%%%%%%%%%%%%%%%%%%%%%%%%

The recent renewed interest in the foundation of quantum statistical mechanics and in the dynamics of isolated quantum systems has led to a revival of the old approach by von Neumann to investigate the problem of thermalization only in terms of quantum dynamics in an isolated system \cite{vonNeumann,GLTZ}.
It has been demonstrated in some general or concrete settings that a pure initial state evolving under quantum dynamics indeed approaches an equilibrium state \cite{Hal1998,Reimann,LindenPopescuShortWinter,
GLMTZ09b,Hal2010,SatoKanamotoKaminishiDeguchi,ReimannKastner,Reimann2}.
The underlying idea that a single pure quantum state can fully describe  thermal equilibrium has also become much more concrete
\cite{PopescuShortWinter,GLTZ06,SugiuraShimizu}.

We must note, however, that in the general theories of the approach to equilibrium 
\cite{vonNeumann,GLTZ,Hal1998,Reimann,LindenPopescuShortWinter,
GLMTZ09b,Hal2010,ReimannKastner,Reimann2}, the issue of the time scale required for the relaxation has not been fully addressed.
Usually a statement about the relaxation is proved for ``sufficiently long (but finite) time'', but no concrete estimates are made of how long the ``finite time'' will be.
Although there is an interesting attempt \cite{ShortFarrelly} to deal with the time-scale, we find their main result not very useful for large systems \cite{Ne}.
If it happens that the required time is as long as, say, the age of the universe, the statement about the approach to equilibrium may not be physically relevant.

In the present paper we prove two theorems for a general class of isolated macroscopic quantum systems.
Although the theorems may look somewhat artificial, they are of direct physical relevance to the above mentioned issue of the time scale as we shall explain below.

Our first theorem is a warning;  it states that there always exists an equilibrium subspace for which the relaxation to equilibrium takes an extraordinarily long time.
Although this is nothing more than a purely theoretical ``existence proof'', it shows that the general theories \cite{vonNeumann,GLTZ,Hal1998,Reimann,LindenPopescuShortWinter,
GLMTZ09b,Hal2010,ReimannKastner,Reimann2} should be supplemented by extra arguments that guarantee the necessary time scale to be sufficiently short.

Our second theorem, which has the opposite character, gives us a hope; it states that there can be an equilibrium subspace for which any initial state approaches equilibrium within a short amount of time.
Although the subspace we shall construct is artificial, it is expected that a realistic equilibrium subspace shares some essential features with our example.

We hope that the present results serve as a basis of future investigation of the fundamental and important problem of the approach to equilibrium \cite{waitingtime}.

\midskip
\para{Setup and background}%
We consider an abstract model for an isolated macroscopic quantum system in a large finite volume $V$. 
A typical example is a system of $N$ particles confined in a box, where $N/V$ is kept constant when $V$ becomes larger.

Let $\hH$ be the Hamiltonian, and denote by $E_j$ and $\kpj$ the eigenvalue and  the normalized eigenstate, respectively, of $\hH$, i.e., $\hH\kpj=E_j\kpj$.
We focus on the energy interval $[U,U+\DU)$, where $\DU$ is small from the macroscopic point of view but is still large enough to contain many energy levels.
It is convenient to relabel the index $j$ so that $E_j\in[U,U+\DU)$ for $j=1,\ldots,D$ (and only for those).
We shall work with the Hilbert space $\calH$ spanned by all $\kpj$ with $j=1,\ldots,D$, which is often called a {\em microcanonical energy shell}\/.
The dimension $D$ of the energy shell typically behaves like  $D\sim e^{aV}$ 
\cite{V,Ruelle} with a constant $a>0$ which is independent of $V$.

To motivate our theorems, let us briefly describe the problem of the approach to equilibrium.
We shall basically follow \cite{GLTZ,GLMTZ09b,Hal2010}, but the discussion applies to other settings.
We recommend \cite{GLTZ} as an accessible exposition.

We first decompose the energy shell $\calH$ into the equilibrium and the nonequilibrium subspaces as $\calH=\Heq\oplus\Hneq$, where any state $\ket{\ph}$ which is close enough to $\Heq$ represents the equilibrium state \cite{eq}.
A state not close to $\Heq$ represents a nonequilibrium state.
Note that neither the set of equilibrium states nor that of nonequilibrium states forms a subspace of $\calH$.
The subspace $\Heq$ occupies most of the energy shell $\calH$ in the sense that the dimension $d_\mathrm{neq}$ of the nonequilibrium subspace $\Hneq$ satisfies $d_\mathrm{neq}\ll D$.
One then easily finds that {\em a typical state in the energy shell represents the equilibrium state}\/ \cite{typical,exception}.

The next question is whether the approach to equilibrium can be understood from the quantum dynamics.
We start from a normalized initial state $\kpz$ which may not be in $\Heq$, and ask  whether its time evolution
\eq
\ket{\ph(t)}=e^{-i\hH t}\kpz
\en
comes and stays, for most $t$, very close to $\Heq$ when $t$ is large.

In some settings (and under suitable assumptions), one can prove that \cite{Hal1998,Reimann,LindenPopescuShortWinter,
GLMTZ09b,Hal2010,ReimannKastner,Reimann2}
\eq
\frac{1}{T}\int_0^Tdt\,\sbkt{\ph(t)|\hP_{\Hneq}|\ph(t)}\ll1
\lb{approach}
\en
for sufficiently large $T$.
Here and in the following $\hP_{\calH'}$ denotes the orthogonal projection onto a subspace $\calH'$ of $\calH$.
The bound \rlb{approach} implies that, within the time interval $[0,T]$,  the state $\ket{\ph(t)}$ spends most of the time in the close vicinity of the  equilibrium subspace $\Heq$.
This establishes the desired approach to equilibrium (apart from the issue of the time scale).

The bound  \rlb{approach} is established for an arbitrary initial state $\ph(0)\in\calH$ in some works \cite{GLMTZ09b,Hal2010}, and for an arbitrary $\ph(0)$ satisfying certain conditions in other works \cite{Hal1998,Reimann,LindenPopescuShortWinter,ReimannKastner,Reimann2}.
Let us note that, in order to account for the approach to equilibrium in real system, it is probably sufficient to establish a relation like \rlb{approach} for the set of physically realizable initial states (at least those that typically arise), which may be much smaller than the whole $\calH$.

Needless to say, the ``sufficiently large'' $T$ associated with the bound  \rlb{approach} should not be too long if the bound is to be physically meaningful.
This question of time scale is the main issue of the present paper.

\midskip
\para{Main theorems}%
Let us state our first theorem.
It shows that, at least theoretically speaking, the required time scale can be extraordinarily long.

\para{Theorem 1}%
For any dimension $d$ with $0<d\le D$ and any state $\ket{\eta}\in\calH$, there exists a $d$-dimensional subspace $\calH_1\ni\ket{\eta}$  such that 
for any normalized initial state $\kpz\in\calH_1$ one has \cite{3pi}
\eq
\frac{1}{T}\int_0^Tdt\,\sbkt{\ph(t)|\hP_{\calH_1}|\ph(t)}\ge\frac{3}{\pi}>0.95,
\lb{slow}
\en
for any $T$ with
\eq
0<T\le\frac{\pi}{6\DU}d.
\lb{longT}
\en

Suppose that the dimension $d$ is of order $e^{bV}\ll D\sim e^{aV}$ with $0<b<a$, which is typical for $\Hneq$.
Then the right-hand side of \rlb{longT} can easily become  much longer than the age of the universe, and \rlb{slow} shows that $\ket{\ph(t)}$ cannot get far from the subspace $\calH_1$ within this time scale.

Let us make two crucial remarks about the choice of $\calH_1$.
For these we assume that $d\ll D$ and the $E_j$ are nondegenerate.
We first note that $\calH_1$ can be chosen so that any $\kpz\in\calH_1$ satisfies the conditions required for initial states in \cite{Hal1998,Reimann,LindenPopescuShortWinter,ReimannKastner,Reimann2}.
See the beginning of the proof.
We also note that $\calH_1$ can be chosen so that for any $\kpz\in\calH$ one has $T^{-1}\int_0^Tdt\,\sbkt{\ph(t)|\hP_{\calH_1}|\ph(t)}\ll1$ for sufficiently large $T$.
(See ``Energy eigenstate thermalization for $\calH_1$'' below for precise statements.)
But \rlb{longT} shows  that if $\kpz\in\calH_1$ such $T$ should be at least of $O(d/\DU)$, which can be extraordinarily large.

We are {\em not}\/ arguing that the nonequilibrium subspace $\Hneq$ generally shares this disappointing property with $\calH_1$.
We note that our $\calH_1$ is intentionally constructed so as to ``trap'' its elements.

Nevertheless the theorem establishes that there {\em always}\/ exists a subspace which leads to an unphysically long relaxation time.
This makes explicit that the previous general results \cite{vonNeumann,GLTZ,Hal1998,Reimann,LindenPopescuShortWinter,
GLMTZ09b,Hal2010,ReimannKastner,Reimann2} on the approach to equilibrium, although being mathematically rigorous, are physically incomplete; they must be supplemented by extra arguments which guarantee that the required time scale is within the physically acceptable range.
To find such arguments, either for general systems or for specific systems, is a difficult but fascinating challenge in fundamental physics.

The second theorem, which has exactly the opposite character from the first, perhaps provides a hint for such an exploration.
It states that there always exists a subspace with a large dimension which is rarely visited and from which it is easy to reasonably quickly escape.

To state the theorem we assume that there is a constant $c>0$ independent of $V$ such that the density of states within $[U,U+\DU)$ is at least $e^{cV}$ \cite{DOS}.

\para{Theorem 2}%
For any $T_0$ such that $0<T_0\DU\le\{\DU/(2\epsilon_0)\}^2$  (where $\epsilon_0$ is defined in \cite{DOS}), there exists a subspace $\calH_2$, whose dimension is $d\sim e^{cV}$, such that
for any normalized initial state $\kpz\in\calH$ and any $T>0$, one has
\eq
\frac{1}{T}\int_0^Tdt\,\sbkt{\ph(t)|\hP_{\calH_2}|\ph(t)}
\le\frac{2}{\sqrt{T_0\DU}}\Bigl(1+\frac{2T_0}{T}\Bigr).
\lb{fast}
\en
We shall choose the constant $T_0$ so that $T_0\DU\gg1$.

Note that the right-hand side of \rlb{fast} becomes small enough within a  short time \cite{timescale}.
We stress that the bound \rlb{fast} is valid for {\em any}\/ choice of the initial state.
The theorem makes it clear that there can be a situation in which the scenario of the approach to equilibrium in an isolated quantum system works ideally.

Unfortunately our $\calH_2$ is constructed in a highly artificial manner, and there is no chance that it coincides with a realistic nonequilibrium subspace $\Hneq$.
What one would like to show is that in some cases the subspace $\Hneq$ shares  essential features with $\calH_2$, in the sense that approach to equilibrium occurs on a reasonable time scale for most initial nonequilibrium states, with ``most" understood in a sufficiently robust manner.

\midskip
\para{Preparations for the proof}%
We expand the initial state as $\kpz=\sumj\al_j\kpj$, where the coefficients satisfy $\sumj|\al_j|^2=1$.
Then the state at time $t\ge0$ is
\eq
\ket{\ph(t)}=e^{-i\hH t}\kpz=\sumj\al_j\,e^{-iE_jt}\kpj.
\lb{pht}
\en

Both $\calH_1$ and $\calH_2$ will be explicitly constructed below as the subspace spanned by mutually orthogonal normalized states $\kx$ with $\nu=1,\ldots,d$.
By expanding the basis states as $\kx=\sumj\bn_j\kpj$, the overlap with \rlb{pht} is written as $\bkt{\xi^{(\nu)}|\ph(t)}=\sumj(\bn_j)^*\al_j\,e^{-iE_jt}$.

Let $\hP=\sumn\kx\langle\xi^{(\nu)}|$ be the projection onto the subspace spanned by $\kx$ with $\nu=1,\ldots,d$.
This corresponds to $\hP_{\calH_1}$ or $\hP_{\calH_2}$.
Then the time averages in \rlb{slow} or \rlb{fast} is evaluated as
\eqa
\frac{1}{T}&\int_0^T\hspace{-4pt}dt\,
\sbkt{\ph(t)|\hP|\ph(t)}
=
\frac{1}{T}\int_0^T\hspace{-4pt}dt
\sumn\sbkt{\ph(t)|\xi^{(\nu)}}\sbkt{\xi^{(\nu)}|\ph(t)}
\nl
&=
\frac{1}{T}\int_0^T\hspace{-4pt}dt
\sumn\sum_{j,k=1}^D\bn_j(\bn_k)^*\al_j^*\al_k\,e^{i(E_j-E_k)t}
\nl
&=\calD+\calO
\lb{main}
\ena
with the diagonal part
\eq
\calD=\sumn\sumj|\bn_j|^2|\al_j|^2,
\lb{D}
\en
and the off-diagonal part
\eq
\calO=\sumn\sumtwo{j,k=1}{(j\ne k)}^D\bn_j(\bn_k)^*\al_j^*\al_k
\frac{e^{i(E_j-E_k)T}-1}{i(E_j-E_k)T}.
\lb{O}
\en

Finally we decompose the interval $[U,U+\DU)$ into $L$ subintervals with the common width $\DU/L$.
For each $\ell=1,\ldots,L$, we denote by $I_\ell$ the set of $j$ such that $E_j$ is in the $\ell$-th interval $[U+(\DU/L)(\ell-1),U+(\DU/L)\ell)$.

\begin{figure}
\centerline{\epsfig{file=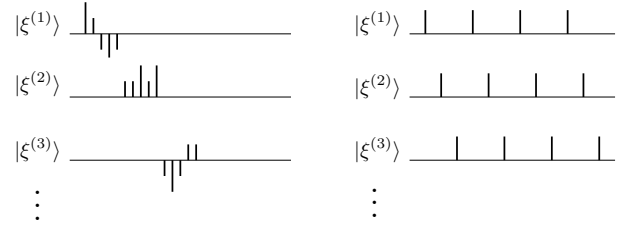,width=8cm}}
\caption[dummy]{
Left:
The basis state $\kx$ of $\calH_1$ is in a narrow energy interval of width $\DU/L$.
Right:
The basis state $\kx$ of $\calH_2$ spreads sparsely over the whole energy range.
}
\label{f:1}
\end{figure}

\midskip
\para{Proof of Theorem 1}%
Here we take $d=L$.
We assume for the moment that no $I_\ell$ is empty.
The only requirement for the basis state $\kx=\sumj\bn_j\kpj$ is that $\bn_j=0$ whenever $j\not\in I_\nu$.
The components of  each basis state are concentrated on a narrow energy interval
as in Fig.~\ref{f:1} (Left).
If $d\ll D$, one can obviously satisfy $|\bn_j|\ll1$, which is essentially the condition required in
\cite{Hal1998,Reimann,LindenPopescuShortWinter,ReimannKastner,Reimann2}.

It is easy to see that $\kx$ can be chosen so that the subspace $\calH_1$ contains an arbitrary given state $\ket{\eta}=\sumj\gamma_j\kpj$.
For each $\nu$ with $\sum_{j\in I_\nu}|\gamma_j|^2\ne0$, we set 
\eq
\bn_j=\begin{cases}
(\sum_{j'\in I_\nu}|\gamma_{j'}|^2)^{-1/2}\,\gamma_j,&j\in I_{\nu};\\
0,&j\ne I_\nu.
\end{cases}
\en
For $\nu$ with $\sum_{j\in I_\nu}|\gamma_j|^2=0$, we choose an arbitrary $\kx$ which satisfies the previously mentioned requirement.

Let us write the initial state as $\kpz=\sumn\zeta_\nu\kx$ with $\sumn|\zeta_\nu|^2=1$.
The coefficient in the expansion $\kpz=\sumj\al_j\kpj$ is $\al_j=\zeta_\nu\bn_j$ where $\nu$ is  such that $j\in I_\nu$.
Then we see that
\eqa
\calO&=\sumn\sumtwo{j,k\in I_\nu}{(j\ne k)}
|\zeta_\nu|^2|\bn_j|^2|\bn_k|^2\,
\frac{e^{i(E_j-E_k)T}-1}{i(E_j-E_k)T}
\nl&
=\sumn\sumtwo{j,k\in I_\nu}{(j\ne k)}
|\zeta_\nu|^2|\bn_j|^2|\bn_k|^2\,
\frac{\sin[(E_j-E_k)T]}{(E_j-E_k)T},
\lb{2nd}
\ena
where we made use of the symmetry between $j$ and $k$, and replaced the summand with its real part.
Note that the function $\sin x/x$ is even, and decreasing for $x\in[0,\pi]$.
Suppose then that $T\DU/L<\pi$.
For any $j,k\in I_\nu$, one has $|E_j-E_k|\le\DU/L$, and hence 
\eq
\frac{\sin[(E_j-E_k)T]}{(E_j-E_k)T}\ge\frac{\sin[T\DU/L]}{T\DU/L}=R,
\lb{R}
\en
with $0<R<1$.
Also note that
\eq
\calD=\sumn\sum_{j\in I_\nu}|\zeta_\nu|^2|\bn_j|^4
\ge R\sumn\sum_{j\in I_\nu}|\zeta_\nu|^2|\bn_j|^4.
\lb{1st}
\en
By substituting \rlb{2nd} and \rlb{1st} into \rlb{main}, we find
\eqa
&\frac{1}{T}\int_0^T\hspace{-4pt}dt\,
\sbkt{\ph(t)|\hP|\ph(t)}
\nl&
\ge
R\sumn\biggl\{\sum_{j\in I_\nu}|\zeta_\nu|^2|\bn_j|^4+
\sumtwo{j,k\in I_\nu}{(j\ne k)}
|\zeta_\nu|^2|\bn_j|^2|\bn_k|^2\biggr\}
\nl&
=R\sumn\sum_{j,k\in I_\nu}|\zeta_\nu|^2|\bn_j|^2|\bn_k|^2
=R.
\ena
To get Theorem~1, we note that the condition \rlb{longT} with $d=L$ implies $T\DU/L\le\pi/6$, and we thus can take $R=\sin(\pi/6)/(\pi/6)=3/\pi$.

It is trivial to remove the assumption that all $I_\nu$ are non-empty.
If one of the $I_\nu$ happens to be empty, we simply split a different $I_{\nu'}$ (with $|I_{\nu'}|\ge2$) into two sets in an arbitrary manner, and repeat the above construction.

\midskip
\para{Energy eigenstate thermalization for $\calH_1$}%
The condition that $\sbkt{\psi_j|\hP_{\calH_1}|\psi_j}\ll1$ for any $j$ may be called the ``energy eigenstate thermalization'' for the subspace $\calH_1$.
It is easily found (see, e.g., \cite{GLMTZ09b}) that this condition and the nondegeneracy of the energy eigenvalues $E_j$ imply
\eq
\frac{1}{T}\int_0^Tdt\,\sbkt{\ph(t)|\hP_{\calH_1}|\ph(t)}\ll1
\lb{H1escape}
\en
when $T$ is sufficiently large for any initial state $\kpz$.
This means that the state gets far from $\calH_1$ in the (very) long run, spending most of its time near the complementary ``equilibrium'' subspace.

Note that $\sbkt{\psi_j|\hP_{\calH_1}|\psi_j}=|\bn_j|^2$, where $\nu$ is  such that $j\in I_\nu$.
One can choose $|\bn_j|^2$ to be extremely small for all $j$ provided that $|I_\nu|\gg1$.
In this way we can construct examples of $\calH_1$ for which the ``approach to equilibrium'' type statement \rlb{H1escape} is valid, but only for $T$ which is extraordinarily large.

\midskip
\para{Proof of Theorem 2}%
Here we shall define completely different basis states $\kx$.
Let $L$ be even and satisfy $\DU/L\ge\epsilon_0$.
For $\nu=1,\ldots,d$, and even $\ell=2,\ldots,L$, we choose $j(\nu,\ell)\in I_\ell$ in such a way that $j(\nu,\ell)\ne j(\nu',\ell)$ if $\nu\ne\nu'$.
This is only possible if $d\le|I_\ell|$ for all $\ell=2,\ldots,L$.
We thus choose $d=\min_\ell|I_\ell|\ge(\DU/L)e^{cV}$.

Then we define the basis states by
\eq
\bn_j=\begin{cases}
\sqrt{2/L},&\text{if $j=j(\nu,\ell)$ for some $\ell=2,\ldots,L$};\\
0,&\text{otherwise}.
\end{cases}
\lb{bn2}
\en
Thus the corresponding basis state $\kx$ spreads sparsely over the whole energy interval $[U,U+\DU)$ as in Fig.~\ref{f:1} (Right).

Then \rlb{D} is bounded as
\eq
\calD=\frac{2}{L}\sumn\sumtwo{\ell=2}{(\text{even})}^L|\al_{j(\nu,\ell)}|^2
\le\frac{2}{L}\sumj|\al_j|^2\le\frac{2}{L}.
\lb{1st2}
\en
To bound \rlb{O}, we note that $\bn_j(\bn_k)^*$ can be nonvanishing only when $j=k$ or $|E_j-E_k|\ge\DU/L$.
Thus, in \rlb{O}, we have
\eq
\abs{\frac{e^{i(E_j-E_k)T}-1}{i(E_j-E_k)T}}\le
\frac{2}{|E_j-E_k|T}\le
\frac{2L}{T\DU}.
\lb{2nd2}
\en
By further noting that 
\eqa
&\sumn\sumtwo{j,k=1}{(j\ne k)}^D
\bigl|\bn_j(\bn_k)^*\al_j^*\al_k\bigr|
%\nl&
\le\frac{2}{L}\sumn\sumtwo{\ell,\ell'=2}{(\ell\ne\ell')}^L
|\al_{j(\nu,\ell)}|\,|\al_{j(\nu,\ell')}|
\nl&
\le\sumn\sumtwo{\ell=2}{(\text{even})}^L|\al_{j(\nu,\ell)}|^2
\le1,
\lb{2nd3}
\ena
where we used the inequality $|\alpha||\alpha'|\le(|\alpha|^2+|\alpha'|^2)/2$, \rlb{O} can be bounded from above by $2L/(T\DU)$.

Substituting this upper bound and \rlb{1st2} into \rlb{main}, we get \cite{logimprovment}
\eq
\frac{1}{T}\int_0^T\hspace{-4pt}dt\,
\sbkt{\ph(t)|\hP|\ph(t)}
\le\frac{2}{L}+\frac{2L}{T\DU}.
\lb{finalH2}
\en
Since \rlb{fast} is trivially satisfied if $T_0\DU\le4$, we assume $T_0\DU\ge4$.
Let us choose $L$ to be the smallest even number greater than or equal to $\sqrt{T_0\DU}$.
Noting that $\sqrt{T_0\DU}\le L\le2\sqrt{T_0\DU}$, we get the desired bound \rlb{fast}. 
The assumption $0<T_0\DU\le\{\DU/(2\epsilon_0)\}^2$ guarantees the condition $\DU/L\ge\epsilon_0$.

\midskip
\para{Discussion}%
We have presented complete proofs of the two theorems by explicitly constructing subspaces of the energy shell $\calH$.
Although our construction is quite artificial, the two subspaces may be regarded as the representatives of the pathological scenario and the ideal scenario in the approach to equilibrium of an isolated large quantum system.

The ``pathological subspace'' $\calH_1$ is spanned by basis states $\kx$ whose components are confined in very narrow energy sub-intervals (see Fig.~\ref{f:1}).
Recalling that each energy eigenstate evolves in time as $e^{-iE_jt}\kpj$, this means that the overlap $|\sbkt{\xi^{(\nu)},\ph(t)}|^2$ changes very slowly.
This is the basic mechanism of the ``trap''.

The ``ideal subspace'' $\calH_2$, on the other hand, is spanned by the basis states $\kx$ whose components are distributed sparsely over the whole energy range (see Fig.~\ref{f:1}).
This construction makes the overlap $|\sbkt{\xi^{(\nu)},\ph(t)}|^2$ vary quickly in time, producing the effective relaxation.

What about more realistic nonequilibrium subspaces? 
We would of course like to show that they are like $\calH_2$, in the sense that all (or at least most) initial states in them relax to equilibrium in a reasonable amount of time. One way of accomplishing this would be to find a useful condition on a subspace that guarantees that this relaxation occurs, and then to show that realistic nonequilibrium subspaces satisfy this sufficient condition.

One might also want to show that most subspaces satisfy this condition. This would be interesting even if one could not show that realistic nonequilibrium subspaces do.

Finally, even without a useful sufficient condition, it would be of value to establish that for most nonequilibrium subspaces this relaxation occurs. Of course, it might be hard to imagine how one could establish this sort of behavior for typical subspaces without having found a useful sufficient condition for the behavior. Nonetheless, numerical simulations could well provide some evidence.

\bigskip
It is a pleasure to thank
Takahiro Sagawa
and
Akira Shimizu
for valuable discussions.

%%%%%%%%%%%%%%%%%%%%%%%%%%%%%%%%%%%%%%%%%%%
%%%%%%%%%%%%%%%%%%%%%%%%%%%%%%%%%%%%%%%%%%%

%%%%%%%%%%%%%%%%%%%%%%%%%%%%%%%%%%%%%%
%%%%%%%%%%%%%%%%%%%%%%%%%%%%%%%%%%%%%%
%%%%%%%%%%%%%%%%%%%%%%%%%%%%%%%%%%%%%%

\begin{thebibliography}{10}
\bibitem{Ne}
It is likely that the key quantity $N(\varepsilon)/d_\mathrm{eff}$ in \cite{ShortFarrelly} is of order $e^{\mathrm{const.}V}$ for large $V$.
See also the remark in \cite{ReimannKastner}.
\bibitem{eq}
Consider the simplest situation where one is interested  in the behavior of a single macroscopic quantity $\hat{A}$, whose equilibrium value is $\bar{A}$.
Then one can define $\Heq$ as the subspace spanned by the eigenstates of the nonnegative operator $(\hat{A}-\bar{A})^2$ corresponding to sufficiently small eigenvalues.
\bibitem{exception}
In the simplest example discussed in \cite{eq}, one can show that the subspace  $\Heq$ occupies most of $\calH$ if the microcanonical average of $(\hat{A}-\bar{A})^2$ is small.
If it happens to be the case that $\Heq$ does {\em not}\/ occupy most of $\calH$, then we do not expect the approach to equilibrium to begin with.
\bibitem{3pi}
The choice of the constant $3/\pi$ in \rlb{slow} is by no means unique.
One may replace it with an arbitrary positive constant less than 1 by suitably modifying \rlb{longT}.
\bibitem{DOS}
More precisely, there is a constant $\epsilon_0$ (with $0<\epsilon_0\ll\DU$) independent of $V$ such that for any $[E,E']\subset[U,U+\DU)$ with $E'-E\ge\epsilon_0$, the number of $j$ such that $E_j\in[E,E']$ is not less than $(E'-E)e^{cV}$.
\bibitem{timescale}
If we make the dependence to Planck's constant explicit, the prefactor in the right-hand side of \rlb{fast} becomes $\sqrt{\hbar/(T_0\DU)}$.
Thus even for $\DU\sim10^{-4}\rm\ J$ and $T_0\sim1\rm\ s$, the factor is as small as $10^{-15}$.
The ``relaxation'' in this artificial example is too effective to be physically realistic.
\bibitem{logimprovment}
By noting that $|E_{j(\nu,\ell)}-E_{j(\nu,\ell')}|\ge|\ell-\ell'|\DU/(2L)$, we can make a better estimate which enables us to replace 
$2L/(T\DU)$ in \rlb{finalH2} with $O(\log L)/(T\DU)$.
\end{thebibliography}
\end{document}